\documentstyle[a4,epsfig]{article}
\twocolumn
\title{{\normalsize \bf MEASUREMENT OF THE MUONIUM 1S-2S TRANSITION
FREQUENCY}}
\author{{\normalsize
V.Meyer$^1$,
S.N.Bagayev$^5$,
P.E.G.Baird$^2$,
P.Bakule$^2$,
M.G.Boshier$^4$,
A.Breitr\"uck$^1$,
S.L.Cornish$^2$,}\\
{\normalsize
S.Dychkov$^5$,
G.H.Eaton$^3$,
A.Grossmann$^1$,
D.H\"ubl$^1$,
V.W.Hughes$^6$,
K.Jungmann$^1$,
I.C.Lane$^2$,
Yi-Wei Liu$^2$,}\\
{\normalsize
D.Lucas$^2$,
Y.Matyugin$^5$,
J.Merkel$^1$,
G.zuPutlitz$^1$,
I.Reinhard$^1$,
P.G.H.Sandars$^2$,
R.Santra$^1$,
P.Schmidt$^1$,}\\
{\normalsize
C.A.Scott$^3$,
W.T.To\-ner$^3$,
M.Towrie$^3$,
K.Tr\"ager$^1$,
C.Wasser$^1$,
L.Willmann$^1$
and
V.Yakhontov$^1$}}
\date{{\normalsize
$^1$Physikalisches Institut, Universit\"at Heidelberg, Philosophenweg 12,
D-69120 Heidelberg, D\\
$^2$Clarendon Laboratory, University of Oxford, Oxford OX1 3PU, UK\\
$^3$Rutherford Appleton Laboratory, Chilton, Didcot, Oxon OX11 0QX, UK\\
$^4$University of Sussex, Physics Department, Brighton BN1 9QH, UK\\
$^5$ Institute of Laser Physics, Novosibirsk 630090, RU\\
$^6$Gibbs Laboratory, Yale University, New Haven, Connecticut 06520-8121, USA}}
\begin{document}
\maketitle
 
%

A new measurement of the 1S-2S energy splitting of muonium
by Doppler-free two-photon spectroscopy has been performed
at the Rutherford Appleton Laboratory in Chilton, Didcot, UK.
Increa\-sed accuracy is expected compared to a previous experiment [1].
Spectroscopy of this transistion promises an improvement of the muon mass
value.\\

One-electron atoms,
being the most fundamental atomic systems,
provide excellent tests for bound state quantum
electrodynamics (QED) and render the possibility
of highly precise measurements of fundamental constants. 
As the energy levels of the natural hydrogen
isotopes (hydrogen, deuterium and tritium)
and hydrogen-like exotic systems with hadronic nuclei
(e.g. muonic helium, pionium and many others)
are influenced
by the finite size and internal structure of the hadrons,
the interpretation
of highly precise measurements in such systems
is limited by todays insufficient
knowledge of the nuclear size effects.
The hydrogen-like muonium atom
($\mu^+e^-$) consists of two leptons from two different generations [2].
No internal structure is known for leptons
down to dimensions of order 10$^{-18}$~m; therefore muonium is free
from nuclear structure effects.
The level energies can be calculated to very high accuracy
exclusively by the theory of bound state Quantum Electrodynamics (QED).
The potential for high precision studies has been
demonstrated in  
a long series of
microwave measurements and theoretical calculations of
the ground state hyperfine structure splitting [2], from which
accurate values for
fundamental constants (muon mass m$_{\mu}$ and
fine structure constant $\alpha$) were obtained [2].
The optical 1S-2S transition offers a
higher resolution than the ground state hyperfine structure splitting,
because of the much
higher transition
frequencies (and QED contributions) and the same $144$~kHz narrow natural linewidth,
which is due to the muon lifetime $\tau_{\mu}$$\approx$2.2$\mu$sec.\\

\begin{figure}[hbt]
\unitlength 1cm
 \begin{picture}(5,5.24)
  \epsfig{file=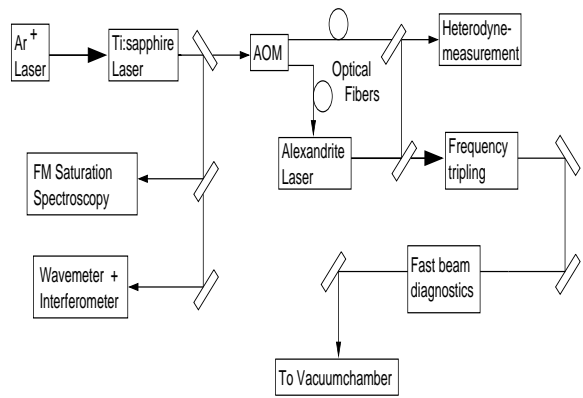,
          height=5.2cm,width=7.6cm}
 \end{picture}\par
 \caption[]{Laser system employed in the experiment.}
\end{figure}

\indent
This experiment was performed at
the worlds brightest pulsed surface muon source
at the
Rutherford Appleton Laboratory (RAL) in Chilton, UK.
The 1$^2$S$_{1/2}$(F=1) $\rightarrow$ 2$^2$S$_{1/2}$(F=1)
transition was induced by Doppler-free two-photon laser spectro\-sco\-py
using two counter-propagating laser beams of wavelength $\lambda = 244$~nm [1].
The atoms were formed
by electron capture after
stopping po\-si\-ti\-ve muons close to the surface of
a SiO$_2$ powder target. A fraction of these
diffused to the surface and left the
powder at thermal velocities (7.43(2)~mm/$\mu$s)
for the adjacent vacuum region.

\indent
The necessary high power UV laser light
was generated
by frequency tripling the output of an alexandrite ring laser amplifier
in crystals of LBO and BBO.
Typically UV light pulses of energy 3~mJ and 80~nsec (FWHM) duration were used.
The alexandrite laser was seeded with light from a continuous wave
Ti:sapphire laser at 732~nm which was pumped by an Ar ion laser.
Fluctuations of the optical phase during the laser pulse were compensated with
an electro-optic device in the resonator of the ring amplifier to give
a frequency chirping of the laser light of less than about 5~MHz.
The laser frequency was calibrated by frequency modulation saturation
spectroscopy of a hyperfine component of the 5-13 R(26) line in thermally
excited iodine vapour.
The frequency of the reference line is about 700~MHz lower than 1/6 of the
muonium transition frequency.
The cw light was frequency up-shifted by passing through two
acousto-optic modulators (AOM's).
The muonium reference line has been calibrated preliminarily to 3.4~MHz at the
Institute of Laser Physics in Novosibirsk. 
An independent calibration at the National Physics Laboratory (NPL) at
Teddington, UK is under way.

\indent
The 1S-2S transition was detected by the
photoionization of the 2S state
by a third photon from the same laser field. The slow muon
set free
in the ionization process is accelerated to 2~keV and guided through a momentum
and energy selective path onto a microchannel plate particle detector (MCP).
Background due to scattered photons and other ionized particles
can be reduced by requi\-ring that the MCP count falls into a 100~nsec wide
window centered at the expected time of flight for mu\-ons
and by additionally requi\-ring
the observation of the energetic positron from the muon decay.
On resonance an event rate of 9 per hour was observed.

\indent
The line shape distortions due to frequency chir\-ping were investigated
theoretically [3,4] and experimentally by observing resonances in deuterium
and hydrogen in the same experimental setup.
A careful analysis is in progress.\\

\begin{figure}[thb]
\unitlength 1cm
 \begin{picture}(5,5.65)
  \epsfig{file=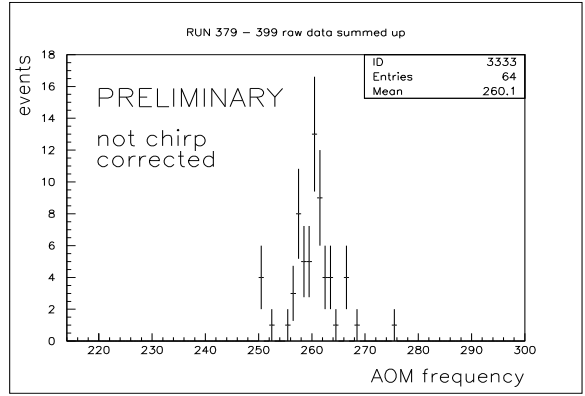,
  height=5.2cm,width=7.6cm}
 \end{picture}\par
 \caption[]{The 
 1$^2$S$_{1/2}$(F=1)$\rightarrow$2$^2$S$_{1/2}$(F=1)
 transition signal in muonium, not corrected for sys\-te\-ma\-tic shifts due to
 frequency chirping and ac Stark effect.}
\end{figure}

This work was supported by the BMBF (Germany), the EPSERC (United Kingdom),
the DOE and NSF (United States), by INTAS and NATO through research grants.

\setlength{\itemindent}{-5mm}
\itemindent -5mm
\setlength{\labelsep}{5pt}
\labelsep 5pt
\parsep 0pt
\parskip 0pt
\itemsep 0pt
\begin{center} {\bf References} \end{center}
\begin{itemize}
\item[{[1]}]F.E. Maas, B. Braun, H. Geerds, K. Jungmann, B.E. Matthias,
 G. zu Putlitz, I. Reinhard, W. Schwarz, L. Willmann, L. Zhang, P.E.G. Baird,
 P.G.H. Sandars, G.S. Woodman, G.H. Eaton, P. Matousek, W.T. Toner, M. Towrie,
 J.R.M. Barr, A.I. Ferguson, M.A. Persaud, E. Riis, D. Berkeland, M.G. Boshier,
 and V.W. Hughes,
 "A measurement of the 1S-2S transition frequency in muonium",\\
 Phys.Lett. A{\bf187}, 247 (1994)
\item[{[2]}]
 V.W. Hughes and G. zu Putlitz,
 {\underline{Quantum} \underline{Electrodynamics}}, T. Kinoshita (ed.),
 "Muonium",
 World Scientific, Singapore, p. 822. (1990)
 and V.W. Hughes,
 {\underline{Atomic} \underline{Physics}\\ \underline{Methods} \underline{in}
 \underline{Modern} \underline{Research}},
 K. Jungmann, J. Kowalski, I. Reinhard, F. Tr\"ager (ed.)\\
 "High Precision Atomic Spectroscopy of Muonium and Simple Muonic Atoms"
 Springer, Heidelberg p. 21 (1997) 
%
%
\item[{[3]}]V. Yakhontov and K. Jungmann,
 "Light-shift calculation in the ns-states of hydrogenic systems",
 Z.Phys.D{\bf38}, 141 (1996)
\item[{[4]}]V.L. Yakhontov, R. Santra and K. Jungmann,
 "3-photon ionization of hydrogenic atoms by non-monochromatic laser field",
 europhysics conference abstracts {\bf21C}, 356 (1997)
\end{itemize}
\end{document}